\documentclass[conference]{IEEEtran}
\usepackage{graphicx}
\usepackage{booktabs}
\usepackage{amsmath}
\usepackage[hidelinks]{hyperref}
\usepackage{caption}
\usepackage{cite}
\begin{document}

\title{Female YouTube Content Creators in Bangladesh: A Data-Driven Study of Audience and Sentiment}

\author{
\begin{tabular}{cc}
\begin{minipage}[t]{0.45\textwidth}\centering\normalsize
Sumaiya Islam\\[3pt]
\small Institute of Information Technology\\
University of Dhaka\\Dhaka, Bangladesh\\bsse1446@iit.du.ac.bd
\end{minipage} &
\begin{minipage}[t]{0.45\textwidth}\centering\normalsize
Sejuti Sharmin Siddiqui\\[3pt]
\small Institute of Information Technology\\
University of Dhaka\\Dhaka, Bangladesh\\bsse1420@iit.du.ac.bd
\end{minipage} \\[1.2em]
\begin{minipage}[t]{0.45\textwidth}\centering\vspace{10pt}\normalsize
Sumaiya Farhana Islam\\[3pt]
\small Institute of Information Technology\\
University of Dhaka\\Dhaka, Bangladesh\\bsse1450@iit.du.ac.bd
\end{minipage} &
\begin{minipage}[t]{0.45\textwidth}\centering\vspace{10pt}\normalsize
B M Mainul Hossain\\[3pt]
\small Institute of Information Technology\\
University of Dhaka\\Dhaka, Bangladesh\\mainul@iit.du.ac.bd
\end{minipage}
\end{tabular}
}

\maketitle
\thispagestyle{plain}
\pagestyle{plain}

\begin{abstract}
Audience response to female-led YouTube channels in Bangladesh has not been measured at scale, and the comments these channels attract are hard to score because they arrive in three different writing systems. This study examines twelve such channels, two per category across six categories, from January 2021 to August 2026, covering 4{,}717 videos and a census of 558{,}827 top-level comments. No single writing system dominates the comment set: English accounts for 33.5\%, Bangla script for 32.0\%, and Banglish (Bengali typed in Latin script and freely mixed with English) for 24.2\%. Each comment was therefore routed to a model trained on the language it was actually written in rather than translated into English. Two categories, selected for the largest rise and the largest drop in positive sentiment over time, were hand-labeled and used to fine-tune the code-mixed model, raising out-of-fold accuracy from 54.3\% to 78.0\% in Health \& Fitness and from 81.1\% to 85.0\% in Beauty \& Fashion, and moving the first category's reported positive share by nearly eight points. The wider results show diverging publishing trends, highly concentrated audience attention, category-specific sentiment trajectories, and a clean split between event-like and search-driven categories in how quickly audiences respond. The findings underscore that language-aware sentiment analysis is essential for Bangladeshi social media data, that even a language-appropriate general model can be systematically miscalibrated for a specific community, and that measured negative shares are best read as lower bounds.
\end{abstract}


\section{Introduction}
Bangladesh has one of the fastest-growing YouTube audiences in South Asia, and a visible share of that growth is driven by women who build audiences around cooking, makeup, fitness, travel, teaching, and comedy. What is far less visible is how those audiences actually behave: which categories are growing, who leaves the comments, how quickly people react, and whether audience reception has improved or deteriorated over time.

This project addresses those questions empirically. Twelve female-led YouTube channels, two from each of six content categories, were tracked over a five-and-a-half-year window from January 2021 to August 2026. Every video published in that window and every top-level comment on those videos were collected, and each comment was scored for sentiment, giving a picture of how publishing output, audience attention, sentiment, and engagement speed have moved across categories and over time.

A significant technical challenge emerged during the work. Roughly a quarter of all comments are written in Banglish (Bengali typed in Latin script, freely mixed with English), which off-the-shelf sentiment models handle poorly and unevenly across communities. Section IV describes how comments were hand-labeled for two categories to fine-tune a category-specific sentiment model, and Section V shows how much that changed the reported results.

\section{Related Work}

\subsection{Women, YouTube, and Bangladesh's Digital Media}
Bangladesh has a rapidly expanding digital-media audience, making platforms such as YouTube increasingly important spaces for public interaction and audience formation. Recent digital reports estimate approximately 43 million YouTube users in Bangladesh, while research on Bangladeshi women shows that social media can support identity construction, social connection, and self-expression, but these activities remain shaped by gender norms, social monitoring, and expectations about women's online behavior \cite{safa2025,datareportal2026}. Research also suggests that even when women possess digital and media literacy, gender identity and prevailing social norms can constrain their active participation online \cite{tasnima2024}. These findings support treating female-led channels as a distinct research setting rather than as an arbitrary subset of YouTube channels, because audience responses to women may be shaped by gendered expectations and forms of evaluation.

\textbf{Gap.} Existing research establishes the importance of women's digital participation in Bangladesh but provides limited empirical evidence on how audiences respond to female-led YouTube channels at scale, particularly through systematic sentiment analysis of their comments.

\subsection{Bangla, Banglish, and Code-Mixed Sentiment Analysis}
Sentiment analysis of Bangladeshi social-media content is complicated by the coexistence of Bangla script, English, and Banglish. SentiGOLD provides a large, five-class Bangla sentiment benchmark containing 70{,}000 samples, with its best reported model achieving a macro-F1 of 0.62 \cite{islam2023sentigold}. However, standard Bangla benchmarks do not directly address the characteristics of Bengali-English code-mixed text. BnSentMix was developed specifically for this setting, providing 20{,}000 code-mixed Bengali-English samples from sources including Facebook, YouTube, and e-commerce platforms, and demonstrating substantially different performance across sentiment classes and text types \cite{alam2025bnsentmix}. These studies show that language-aware models are necessary, but they primarily establish the difficulty of classification rather than examining how language-specific errors affect estimates of negativity within female-focused online communities.

\textbf{Gap.} Prior work provides suitable resources for Bangla and Banglish sentiment classification but does not establish whether reported negative-sentiment rates in Bangladeshi female-creator communities remain reliable when negative examples are sparse, domain-specific, and linguistically mixed.

\subsection{Online Harassment and Negative Responses Toward Women}
Research on Bangladeshi women's online experiences indicates that negative audience behavior cannot be reduced to ordinary dissatisfaction or negative opinion. Recent analysis of 5{,}929 Facebook comments found widespread misogynistic bullying, aggressive insults, mockery, trolling, and shaming directed at women; 46.57\% of the analyzed comments were classified as misogynistic bullying \cite{hossain2026misogyny}. The study also documented explicitly sexualized and body-focused forms of abuse, demonstrating that hostility toward women can be expressed through language whose meaning depends strongly on gender and context. This is important for sentiment analysis because a comment classified simply as negative may represent very different phenomena, ranging from criticism to misogynistic abuse. Conversely, a low measured negative percentage should not automatically be interpreted as evidence that hostility is rare. In the present study, sentiment is calculated only from top-level comments, while reply-thread interactions are excluded; therefore, hostile exchanges occurring in replies are not represented in the measured negative share.

\textbf{Gap.} Existing research demonstrates substantial gendered hostility online, but little work connects this evidence with large-scale, language-aware sentiment analysis of Bangladeshi female YouTube creators; consequently, a low negative-sentiment percentage in top-level comments should be treated as an observed lower-bound estimate of hostility rather than a settled measure of the community's actual negative behavior.

\section{Dataset}
This dataset contains YouTube comments on the most well-known female content creator channels in Bangladesh, along with video metadata, engagement statistics, comment language, sentiment labels, confidence scores, and anonymized comment text. The major columns include: category, channel\_name, video\_title, year, video\_comment\_count, video\_like\_count, comment\_text, language, sentiment, and confidence\_score.

\subsection{Data Source and Collection}
Twelve channels were selected, two per category, each pair chosen to include one large and one smaller creator so that no single mega-channel could dominate its category's results.

\begin{table}[htbp]
\caption{The twelve channels studied, paired so each category contains one large and one smaller creator.}
\begin{center}
\setlength{\tabcolsep}{3.5pt}
\begin{tabular}{lll}
\toprule
\textbf{Category} & \textbf{Channel} & \textbf{Subscribers} \\
\midrule
Beauty \& Fashion & Shahnaz Shimul & 1.43M \\
Beauty \& Fashion & Makeup Maniac By Linda & 754K \\
Food & Cooking Studio by Umme & 2.63M \\
Food & RB Kitchen & 2.00M \\
Travel & Jannat The Lunatic Traveler & 175K \\
Travel & Shurovy Yeasmin & 51K \\
Health \& Fitness & Tamanna Chowdhury & 109K \\
Health \& Fitness & Coach Kriti (Sumaiya Chowdhury) & 11.1K \\
Comedy \& Ent. & Thoughts of Shams & 840K \\
Comedy \& Ent. & Raba Khan & 286K \\
Education \& Skills & Munzereen Shahid & 2.19M \\
Education \& Skills & Maisun's World & 1.17M \\
\bottomrule
\end{tabular}
\label{tab:channels}
\end{center}
\end{table}

Data was collected through the YouTube Data API v3. For each channel, the collector resolved the uploads playlist, enumerated the channel's full upload history, retained only videos published inside the study window, fetched view/like/comment statistics in batches, and paged through every top-level comment thread with no cap. The comment set is therefore a census of top-level comments, not a sample. The 10{,}000-unit daily API quota required spreading collection across several days with resume support, and a backfill pass corrected an early run that had stored truncated comment text on a subset of videos.

\subsection{Data Description}
Collection produced two raw tables, videos.csv (4{,}717 videos, 13 columns) and comments.csv (558{,}827 unique top-level comments, 8 columns), which were merged with sentiment output into a single analysis file of 558{,}827 rows and 20 columns, one row per comment. Table \ref{tab:composition} shows how volume distributes across categories.

\begin{table}[htbp]
\caption{Dataset composition. Video output and audience volume are close to independent: Comedy \& Entertainment produces 27$\times$ more comments than Travel from only twice as many videos.}
\begin{center}
\begin{tabular}{lrrrr}
\toprule
\textbf{Category} & \textbf{Videos} & \textbf{Videos w/} & \textbf{Comments} & \textbf{\%} \\
 & & \textbf{comments} & & \\
\midrule
Comedy \& Ent. & 793 & 788 & 211,380 & 37.8\% \\
Food & 1,373 & 1,368 & 148,919 & 26.6\% \\
Education \& Skills & 1,170 & 772 & 132,047 & 23.6\% \\
Beauty \& Fashion & 271 & 270 & 51,095 & 9.1\% \\
Health \& Fitness & 719 & 638 & 7,771 & 1.4\% \\
Travel & 391 & 364 & 7,615 & 1.4\% \\
\midrule
Total & 4,717 & 4,200 & 558,827 & 100\% \\
\bottomrule
\end{tabular}
\label{tab:composition}
\end{center}
\end{table}

Structurally, the data required little repair since it originates from an API rather than manual entry: comment IDs are unique by construction, timestamps arrive as ISO-8601, and 408 videos with comments disabled were retained in the video table but contribute no comment rows.

\subsection{How the Audience Actually Writes}
\begin{figure}[!htbp]
\centering
\includegraphics[width=0.95\linewidth]{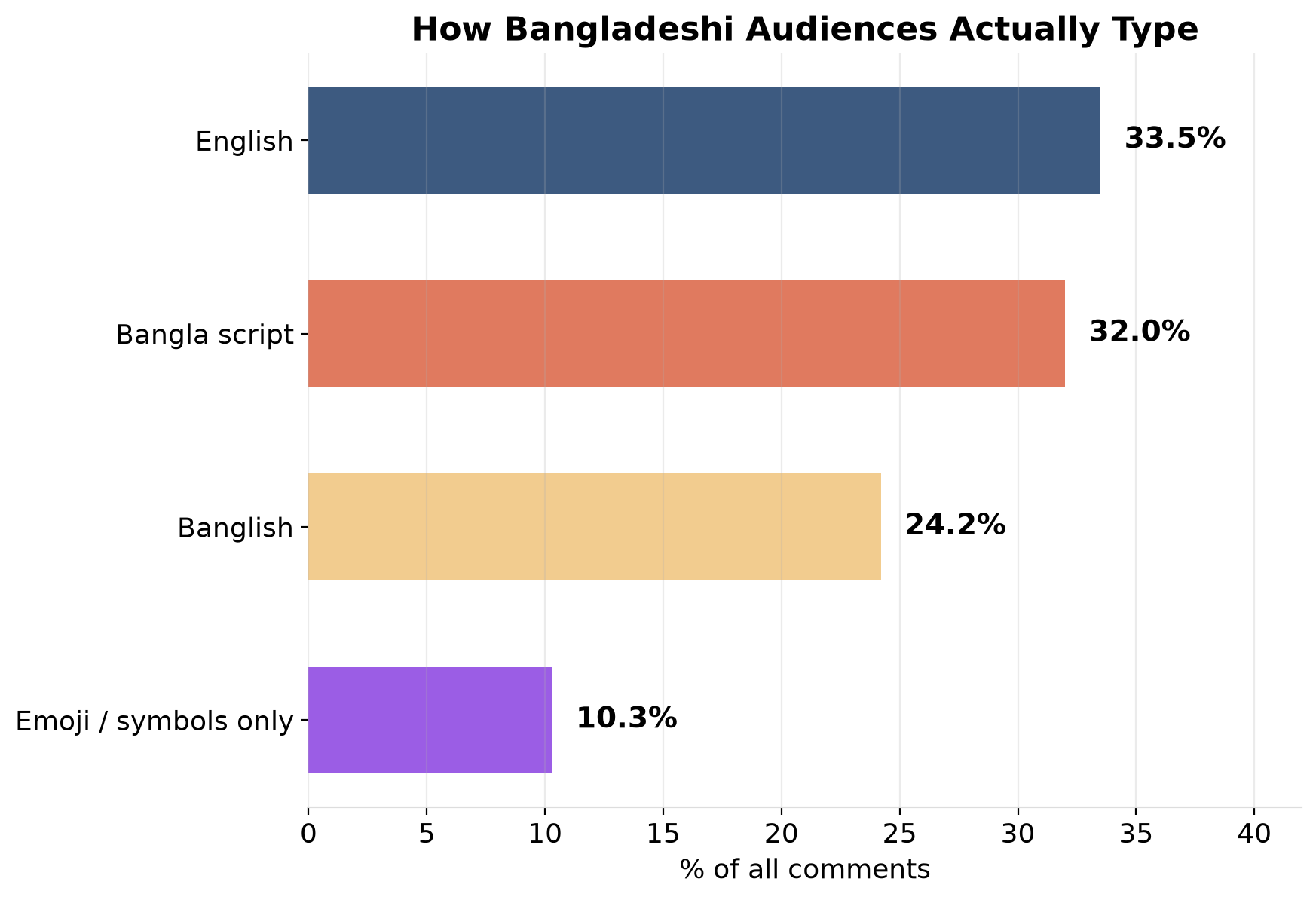}
\caption{Writing systems used across the 558{,}827 collected comments. No single language dominates: English, Bangla script and Banglish each account for a quarter to a third of the set, which is why a single English sentiment tool could not have been used.}
\label{fig:languagemix}
\end{figure}

Before any sentiment work could begin, the writing system of each comment had to be established, because it determines which model can score it. A rule-based classifier assigns each comment to one of four groups. If more than half its letters fall in the Bengali Unicode block it is Bangla. Otherwise, if it contains any of roughly a hundred high-frequency Bengali words commonly typed in Latin script (\textit{apu}, \textit{bhai}, \textit{onek}, \textit{bhalo}, \textit{sundor}, \textit{kemon} and similar) it is Banglish. Remaining Latin text is English. Anything with no letters at all, typically emoji or punctuation, is non-textual. Fig.~\ref{fig:languagemix} and Table~\ref{tab:language} give the result.

\begin{table}[htbp]
\caption{Language composition of the comment set. Emoji-only comments carry scorable sentiment and are retained; the non-textual rows do not and are excluded from every sentiment percentage reported here.}
\begin{center}
\begin{tabular}{lrr}
\toprule
\textbf{Writing system} & \textbf{Comments} & \textbf{Share} \\
\midrule
English & 187,151 & 33.5\% \\
Bangla script & 178,927 & 32.0\% \\
Banglish (code-mixed) & 135,148 & 24.2\% \\
Emoji only & 48,263 & 8.6\% \\
Non-textual (no scorable emoji) & 9,338 & 1.7\% \\
\midrule
Total & 558,827 & 100\% \\
\bottomrule
\end{tabular}
\label{tab:language}
\end{center}
\end{table}

The mix is not uniform across categories, and that turns out to matter a great deal for Section V. Beauty \& Fashion is 51.9\% Banglish, more than double the overall rate, while Education \& Skills is 54.1\% English and Food is 50.6\% Bangla script. A model weakness in Banglish therefore damages the Beauty \& Fashion result far more than it damages the Education result.

\subsection{Data Cleaning and Privacy Processing}
Because the dataset consists of real, identifiable comments from the public, a dedicated privacy-cleaning process was applied before analysis. All comment-author identifiers such as author usernames, user IDs and comment IDs were removed from the working dataset, along with contact-detail and social-profile fields, so that no record can be traced back to an individual account.

The comment text itself was then scanned for embedded personal information. Detected email addresses, phone numbers, URLs, and social-media handles appearing inside comment bodies were replaced with neutral placeholder tokens rather than left in place or simply deleted, preserving sentence structure for downstream sentiment scoring while removing the identifying content. Location fields associated with commenters were removed on the same basis.

As an additional safeguard, a multilingual named-entity (NER) audit was run over the comment text using spaCy to flag potential person, organization, location, and miscellaneous entities. Because a substantial share of comments are Banglish, which general-purpose NER models were not trained on, this step produces a review report of flagged spans rather than automatically deleting or redacting them. Automatic removal risked a high false-positive rate on code-mixed text, silently damaging comment text that contained no real personal information. The audit output was used to manually review high-confidence detections rather than to alter the dataset directly.

\section{Algorithms \& Experiments}
After data collection, preparation and cleaning, the final dataset was subsequently subjected to sentiment analysis using appropriate models based on the comment language. Their outputs were compared to identify differences in sentiment patterns across categories. Following this, two categories, selected based on highest rise and highest drop in positive sentiments after a time series analysis, were fine-tuned with a hand-labelled subset of the dataset. The resulting findings were then interpreted in terms of the observed patterns, implications, limitations, and recommendations.

\begin{figure*}[htbp]
\centering
\includegraphics[width=0.95\textwidth]{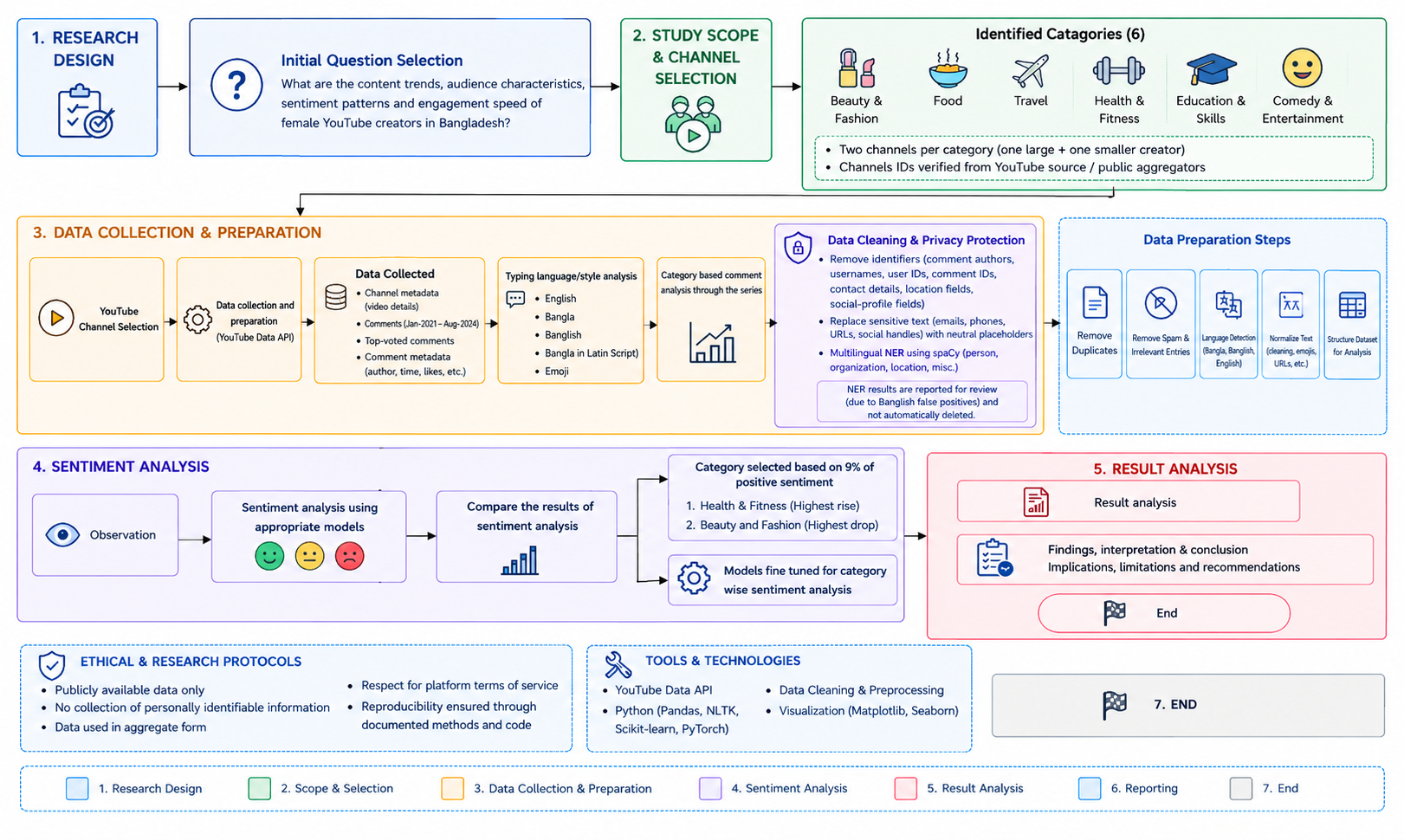}
\caption{The complete methodology diagram.}
\label{fig:methodology}
\end{figure*}

\subsection{Sentiment-Scoring Pipeline}
Rather than translating everything into English and losing meaning in the process, each comment is routed to a model trained on the language it was actually written in, using the classifier from Section III-C. All models emit one of three labels (positive, negative, neutral) with a confidence score, so the outputs can be pooled into a single column.

\begin{itemize}
\item \textbf{Bangla script (32.0\%).} A BanglaBERT \cite{bhattacharjee2022banglabert} multiclass classifier trained on SentiGOLD \cite{islam2023sentigold}. The published checkpoint ships only generic \texttt{LABEL\_0} through \texttt{LABEL\_4} names with no label map, so the mapping was established by probing the model with comments of known polarity before it was trusted on the full set.
\item \textbf{Banglish (24.2\%).} A code-mixed Bengali-English sentiment model built on XLM-RoBERTa \cite{conneau2020xlmr}, trained on BnSentMix \cite{alam2025bnsentmix}. This is the model fine-tuned per category in Section IV-C.
\item \textbf{English (33.5\%).} CardiffNLP's Twitter-RoBERTa sentiment model \cite{barbieri2020tweeteval}, trained on social-media English rather than formal prose, which suits comment text.
\item \textbf{Emoji only (8.6\%).} Comments made entirely of emoji carry real sentiment, and discarding them would have thrown away 48{,}263 records. Each emoji is looked up in the Emoji Sentiment Ranking lexicon \cite{novak2015emoji} and the comment takes the mean score of its emoji. 9{,}338 comments contained no emoji present in the lexicon, usually pure punctuation or symbols, and are marked unscorable and excluded from every sentiment percentage reported here.
\end{itemize}

After scoring, 549{,}489 of 558{,}827 comments carry a usable sentiment label. Across the whole dataset the split is 57.6\% positive, 35.2\% neutral, 6.9\% negative, and 0.3\% mixed, an aggregate that masks substantial variation by category and year, presented in Section V.

\subsection{The Hand-Labeled Data}
A general Banglish model has seen a broad sample of code-mixed Bengali, but it has not seen how a particular community talks. A fitness audience writes long, factual questions about diet plans and routines that read as flat text but are warmly intended. A makeup audience writes short, emphatic compliments. The suspicion was that a single model would systematically misread at least one of these registers, and the way to test that was to hand-label real comments from each category and measure the model against them.

All Banglish comments were extracted for the two selected categories, and a sample from each was labeled manually using three classes only: 0 for positive, 1 for negative and 2 for neutral. The Health \& Fitness set was annotated first, in a spreadsheet, and converted to CSV; the Beauty \& Fashion set followed the same scheme.

Two authors individually annotated 1{,}000 samples from each of the two categories, and cases where they disagreed were resolved by the judgment of the third author. Of each set, 900 comments were used for fine-tuning and the remaining 100 were held out to measure the model's accuracy and F1 score before and after fine-tuning.

The class balance differs sharply between the two sets, and that difference drives most of what follows. Health \& Fitness is 71\% positive with 21 negatives. Beauty \& Fashion is close to an even split between positive and neutral with a reasonable 74 negatives. A model can learn a class it has seen 74 times; it cannot learn much from 21.

\subsection{Building the Model}
The base model is an XLM-RoBERTa classifier with 278 million parameters and a 250{,}002-token vocabulary, and its head predicts four classes: positive, negative, neutral and mixed. The annotation scheme uses the first three of those in the same order. Rather than discard the head and start from a random one, the three matching rows of the output projection were copied into a new three-class layer and the mixed row was dropped. The model therefore starts training already knowing what positive, negative and neutral look like in code-mixed Bengali, which is essential when only a few hundred examples are available to learn from.

A second decision made training possible on the available hardware, a laptop GPU with 4 GB of memory. Of the 278 million parameters, 192 million sit in the embedding table. Every parameter matching the embedding layers was frozen. This cuts memory use to roughly 2.9 GB, which fits, and it also acts as regularization: with the embeddings fixed, a few hundred examples cannot pull the model's vocabulary representation off its pretrained footing.

\subsection{Training Setup and Evaluation Protocol}
Both categories were trained with identical settings so their results are comparable: sequences truncated to 128 tokens, batch size 16, AdamW at a learning rate of 2e-5 with weight decay 0.01, a linear schedule with 10\% warmup, gradient clipping at 1.0, mixed-precision fp16, four epochs, and a fixed seed of 42. Class weights were computed as the square root of the standard balanced weights. Plain balanced weighting would have given the 21-example negative class in Health \& Fitness a weight roughly 16 times that of positive, which destabilises training on a set this small; the square root keeps the correction without the explosion.

Performance was measured with 5-fold stratified cross-validation. Each fold trains a fresh model on 80\% of the 900 fine-tuning comments and predicts the held-out 20\%, and the five sets of held-out predictions are concatenated. Every reported figure is therefore an out-of-fold prediction: no comment was ever scored by a model that had seen it during training. A final model was then trained on all labeled data and used to score the rest of the category.

\subsection{Cross-Validated Results}
The general-model column in Table \ref{tab:finetune} is not an estimate. It is the general model's own existing prediction for those exact comments, taken from the sentiment run in Section IV-A, so the comparison is like for like on identical rows.

\begin{table}[htbp]
\caption{Out-of-fold cross-validation performance. The weaker the general model was in a category, the more fine-tuning recovered.}
\begin{center}
\begin{tabular}{llccc}
\toprule
\textbf{Category} & \textbf{Metric} & \textbf{General} & \textbf{Fine-} & \textbf{Change} \\
 & & \textbf{model} & \textbf{tuned} & \\
\midrule
Health \& Fitness & Accuracy & 54.3\% & 78.0\% & +23.7 pts \\
Health \& Fitness & Macro-F1 & 0.439 & 0.606 & +0.167 \\
Beauty \& Fashion & Accuracy & 81.1\% & 85.0\% & +3.9 pts \\
Beauty \& Fashion & Macro-F1 & 0.733 & 0.761 & +0.028 \\
\bottomrule
\end{tabular}
\label{tab:finetune}
\end{center}
\end{table}

Macro-F1 is reported alongside accuracy because accuracy alone is misleading on imbalanced data. A model that labeled every Health \& Fitness comment positive would score 71\% accuracy while detecting nothing. Macro-F1 averages the three classes equally, so a class the model cannot handle drags the number down no matter how rare it is.

\begin{table}[htbp]
\caption{Per-class F1. Negative is the hardest class for both models, and the only one where fine-tuning barely helps Beauty \& Fashion.}
\begin{center}
\begin{tabular}{lcccc}
\toprule
\textbf{Class} & \textbf{H\&F} & \textbf{H\&F} & \textbf{B\&F} & \textbf{B\&F} \\
 & \textbf{general} & \textbf{fine-tuned} & \textbf{general} & \textbf{fine-tuned} \\
\midrule
Positive & 0.65 & 0.86 & 0.87 & 0.88 \\
Neutral & 0.46 & 0.60 & 0.81 & 0.87 \\
Negative & 0.20 & 0.36 & 0.52 & 0.53 \\
\bottomrule
\end{tabular}
\label{tab:perclass}
\end{center}
\end{table}

\begin{figure*}[t]
\centering
\includegraphics[width=0.60\textwidth]{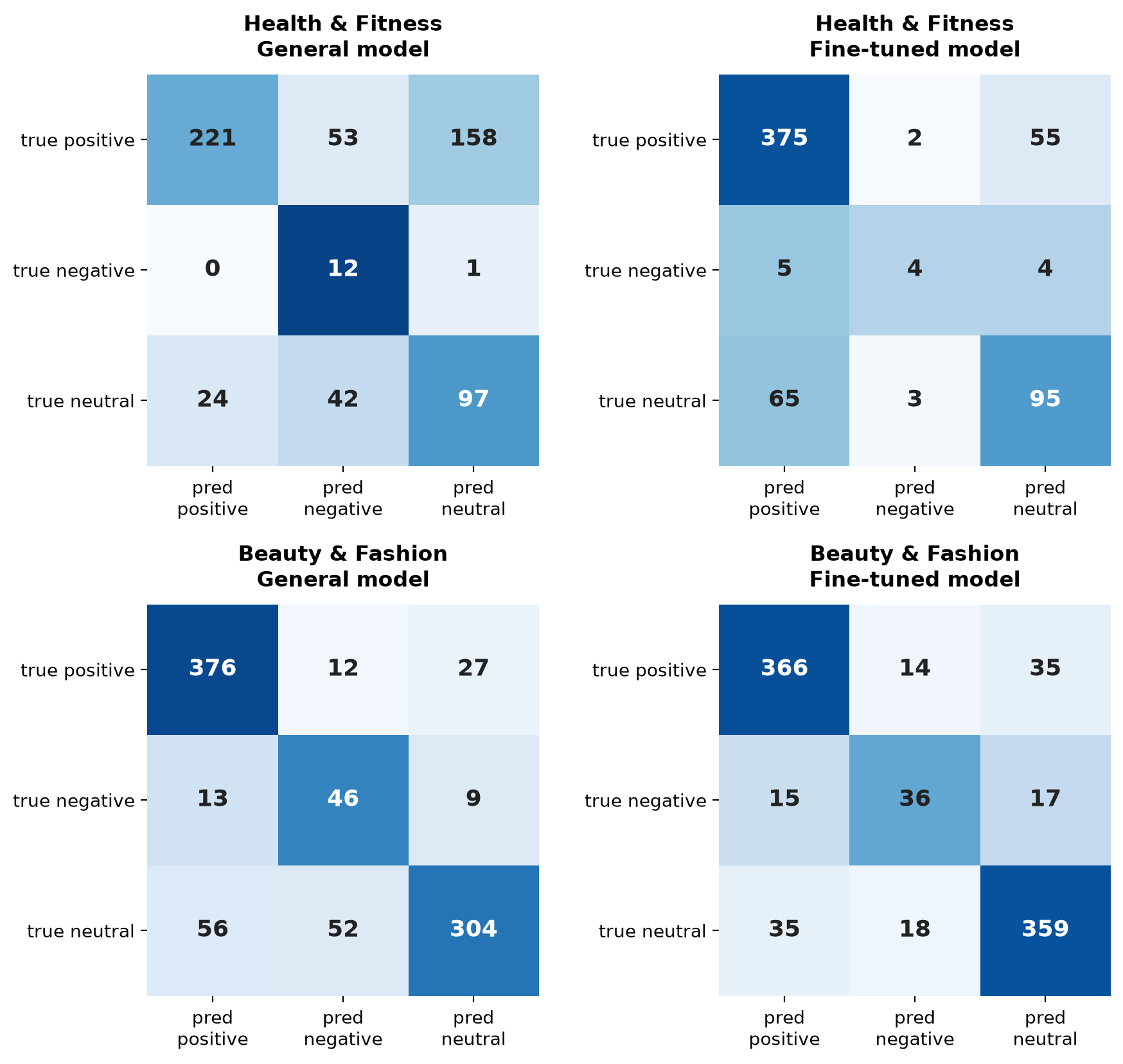}
\caption{Out-of-fold confusion matrices before and after fine-tuning. The Health \& Fitness general model scatters true positives across all three predicted classes, sending 158 of them to neutral and 53 to negative; the fine-tuned model recovers 375 of 430. Beauty \& Fashion starts with a well-formed diagonal, so fine-tuning mainly tightens the neutral row rather than changing the shape of the error.}
\label{fig:confusion}
\end{figure*}

Negative sentiment remained the hardest class for both fine-tuned models, reflecting how few negative examples were available to learn from. Fig.~\ref{fig:confusion} shows where those errors actually sit. Results resting on the exact negative rate in these two categories should be treated as indicative rather than settled.

\section{Result Analysis}

\subsection{Content Output Over Time}
Publishing output tells a different story per category. Travel is the clear growth story, rising from 9 videos in 2021 to 114 in 2026 (a partial year). Education \& Skills roughly doubled its output. Health \& Fitness moved the other way, collapsing from a peak of 215 videos in 2022 to just 14 in 2026, the steepest contraction in the study, while Food remained the largest and most stable producer throughout.

\subsection{Sentiment Trends}
Aggregated across all twelve channels, sentiment is positive and modestly improving, rising from 56.8\% positive on 2021 videos to 62.2\% on 2026 videos, with dips in 2022 and 2024 (2024 also carries the study's highest negative share, 8.4\%).

\begin{figure}[!htbp]
\centering
\includegraphics[width=0.95\linewidth]{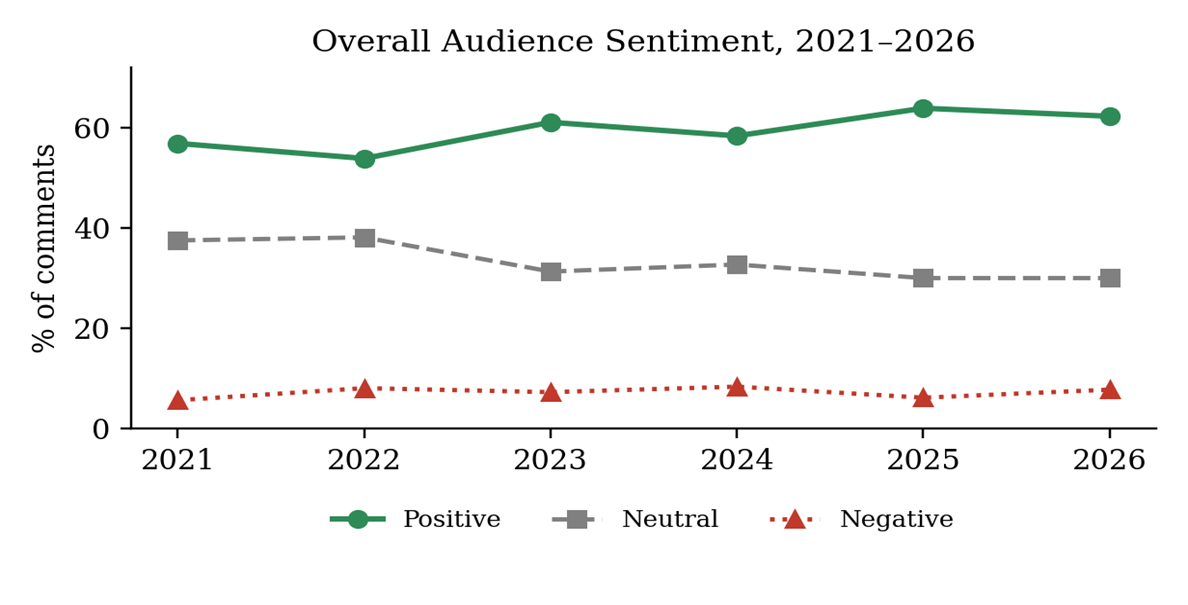}
\caption{Overall sentiment by video publication year across all twelve channels.}
\label{fig:sentimenttrend}
\end{figure}

The category-level picture is more informative than the aggregate (Table \ref{tab:categorysentiment}). Food carries the highest positive share at 65.1\% and is still improving, reaching 70.6\% on 2026 videos. Education \& Skills shows the single largest turnaround in the study, from 35.0\% positive in 2021 to 64.7\% in 2026. Travel is the only category moving decisively the wrong way while its output grows fastest of all, and it carries the highest negative share of any category at 10.0\%. Beauty \& Fashion, despite producing the fastest audience reactions of any category, shows the sharpest decline in positive sentiment.

\begin{table}[htbp]
\caption{Positive sentiment in the first and last year of the study window. The 2026 figures for Health \& Fitness (n=29) and Beauty \& Fashion (n=301) are directional given their small sample sizes.}
\begin{center}
\begin{tabular}{lccc}
\toprule
\textbf{Category} & \textbf{\% pos.} & \textbf{\% pos.} & \textbf{Change} \\
 & \textbf{2021} & \textbf{2026} & \\
\midrule
Health \& Fitness & 51.0\% & 79.3\% & +28.3 pts \\
Education \& Skills & 35.0\% & 64.7\% & +29.7 pts \\
Food & 61.6\% & 70.6\% & +9.0 pts \\
Comedy \& Entertainment & 60.3\% & 58.1\% & $-$2.2 pts \\
Travel & 57.1\% & 45.3\% & $-$11.8 pts \\
Beauty \& Fashion & 62.6\% & 41.9\% & $-$20.7 pts \\
\bottomrule
\end{tabular}
\label{tab:categorysentiment}
\end{center}
\end{table}

\begin{figure}[!htbp]
\centering
\includegraphics[width=0.95\linewidth]{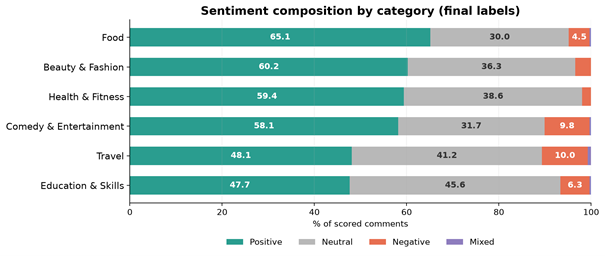}
\caption{Sentiment composition per category using the final labels, including the fine-tuned results for Health \& Fitness and Beauty \& Fashion.}
\label{fig:composition}
\end{figure}

\subsection{Effect of Fine-Tuning on Reported Findings}

\begin{figure}[!htbp]
\centering
\includegraphics[width=0.95\linewidth]{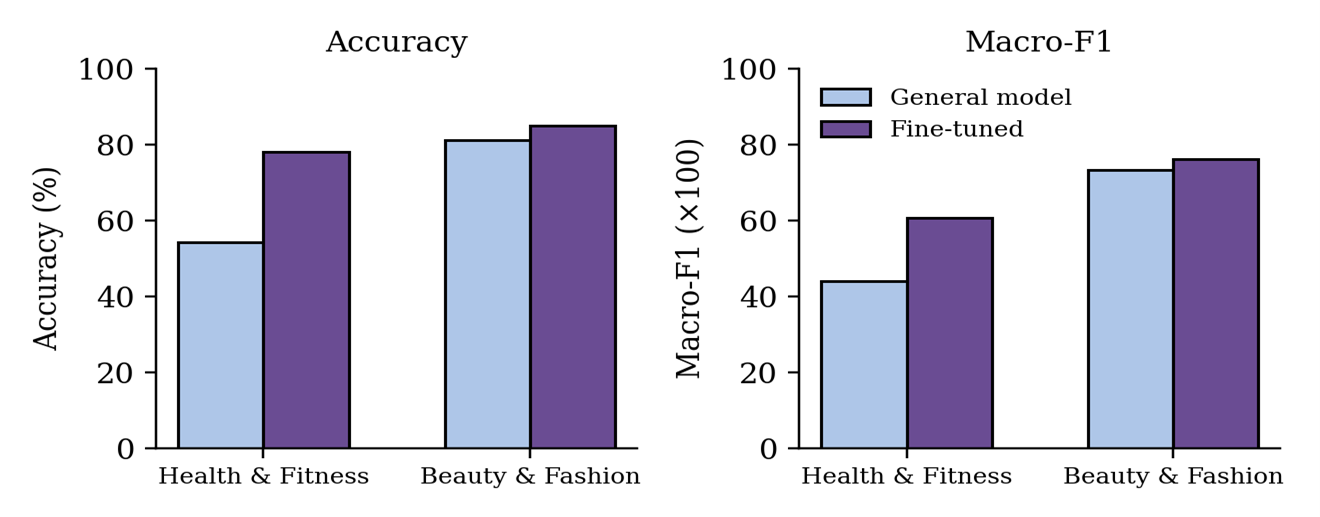}
\caption{Accuracy and macro-F1 before and after fine-tuning. The Health \& Fitness gain is roughly six times the Beauty \& Fashion gain, because the general model was already accurate on Beauty \& Fashion's short, emphatic comment style.}
\label{fig:finetunechart}
\end{figure}

Applying the fine-tuned models changed a meaningful share of Banglish labels: 47.5\% of Health \& Fitness Banglish comments received a different label, against 11.1\% for Beauty \& Fashion. The two corrections point in different directions. In Health \& Fitness the fine-tuned model moves 611 comments into positive, taking them roughly evenly from neutral and negative: within that category's Banglish comments, positive rises from 36.8\% to 71.5\% and negative collapses from 18.1\% to 1.3\%. In Beauty \& Fashion the dominant movement is 2{,}084 comments into neutral, largely from an over-predicted negative class that falls from 8.9\% to 4.5\%.

\begin{figure}[!htbp]
\centering
\includegraphics[width=0.95\linewidth]{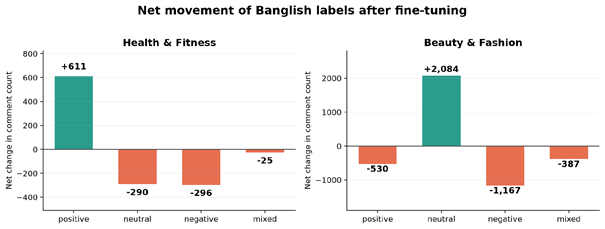}
\caption{Net movement of labels within each category's Banglish comments. Health \& Fitness gains positives at the expense of everything else; Beauty \& Fashion mostly converts over-called negatives into neutrals.}
\label{fig:relabel}
\end{figure}

At the level of the whole category, including the Bangla, English and emoji comments that were untouched, the effect is smaller but still material (Table \ref{tab:categoryeffect}). The master dataset was backed up before the first merge, the aggregate files were backed up before regeneration, and the regeneration method was validated by reproducing the original aggregates byte for byte from the backup before the new ones were trusted.

\begin{table}[htbp]
\caption{Effect on whole-category sentiment, across all languages. The other four categories are unchanged, since no fine-tuned model was applied to them.}
\begin{center}
\begin{tabular}{llcc}
\toprule
\textbf{Category} & \textbf{Sentiment} & \textbf{Before} & \textbf{After} \\
\midrule
Health \& Fitness & Positive & 51.51\% & 59.41\% \\
Health \& Fitness & Neutral & 42.40\% & 38.65\% \\
Health \& Fitness & Negative & 5.77\% & 1.94\% \\
Beauty \& Fashion & Positive & 61.28\% & 60.24\% \\
Beauty \& Fashion & Neutral & 32.22\% & 36.31\% \\
Beauty \& Fashion & Negative & 5.74\% & 3.44\% \\
\bottomrule
\end{tabular}
\label{tab:categoryeffect}
\end{center}
\end{table}

\subsection{Engagement Speed}
Engagement speed, the share of a video's comments arriving within 24 hours of publication, separates categories that behave like events from those that behave like reference material. Beauty \& Fashion and Comedy \& Entertainment are fastest, at 61.8\% and 57.3\% respectively, consumed by audiences waiting for new uploads. Education \& Skills sits at the opposite end, at just 21.4\%, consistent with tutorials being found through search long after upload rather than consumed as they appear; its 132{,}047 comments across 772 videos reflect a long useful life rather than weak interest.

\section{Discussion}

\subsection{Why the Two Fine-Tuned Categories Behaved Differently}
The clearest methodological result of the project is that the value of fine-tuning was inversely proportional to how well the general model already performed. Health \& Fitness gained 23.7 accuracy points; Beauty \& Fashion gained 3.9. The reason is visible in the confusion matrices of Fig.~\ref{fig:confusion}. The general model was misreading an entire register in Health \& Fitness, treating detailed, unemotional questions about diet and training as neutral or hostile, and once shown a few hundred examples it stopped. In Beauty \& Fashion the register is short, emphatic praise, which is exactly what a general code-mixed sentiment model is already good at, so there was little left to gain.

This has a practical implication that generalises beyond this dataset. The decision to hand-label data should not be made uniformly across a corpus. It should be spent where a general model is measurably failing, and the only way to know where that is, is to label a small sample per segment and measure. In this project an identical effort of 1{,}000 labels bought a 23.7 point improvement in one category and only 3.9 points in the other.

It also means the headline sentiment figures for Health \& Fitness differ substantially from what the general model alone would have reported: 59.4\% positive rather than 51.5\%, and 1.9\% negative rather than 5.8\%. The measurement instrument changed the finding.

\subsection{Limitations}
\begin{itemize}
\item \textbf{Sample size.} Twelve channels cannot represent every female creator in Bangladesh. They were chosen for verifiability and category balance, not by random sampling, so these are findings about this cohort.
\item \textbf{Category imbalance.} Health \& Fitness and Travel contribute under 3\% of all comments between them. Their year-level percentages, particularly for 2026, rest on samples in the hundreds or lower and are directional only.
\item \textbf{Uneven sentiment validation.} Sentiment was validated against human labels only for Banglish comments in Health \& Fitness and Beauty \& Fashion. The Bangla-script, English and emoji scorers, and all four remaining categories, rely on unvalidated model output. The Bangla model in particular ships without a label map, and its mapping was established by probing rather than from documentation.
\item \textbf{Negative sentiment detection.} Negative is the hardest class for both fine-tuned models, with F1 of 0.36 in Health \& Fitness and 0.53 in Beauty \& Fashion. The near-zero negative rate now reported for Health \& Fitness Banglish comments almost certainly understates the true rate.
\item \textbf{Top-level comments only.} Replies to comments were not collected. Discussion that happens inside reply threads, which is often where disagreement lives, is absent from this dataset, and its absence probably biases the overall picture positive. Given the evidence of gendered hostility in Bangladeshi online spaces \cite{hossain2026misogyny}, the negative shares reported here are best read as a lower bound rather than a settled measure of the hostility these creators receive.
\end{itemize}

\section{Conclusion}
Publishing output is rising overall and shifting toward Travel and Education \& Skills, while Health \& Fitness contracts sharply; audience attention remains highly concentrated, with three categories holding 88\% of all comments. Sentiment is positive and slowly improving in aggregate, with Food and Education \& Skills strong and improving while Travel and Beauty \& Fashion decline for different reasons: the former growing but souring, the latter fast-reacting but cooling. Response speed splits cleanly between event-like categories and search-driven ones.

The most transferable result of the project concerns measurement itself: language-aware sentiment scoring is not optional for Bangladeshi social media data, where no single writing system accounts for more than a third of comments, and even a language-appropriate general model can be systematically wrong for a specific community. Two modest annotation efforts of 1{,}000 comments each were enough to detect this and to correct it where it mattered most.

Three extensions would strengthen the work. Labeling a few hundred comments in the remaining four categories would establish whether the general model is failing there too, at a cost of a few hours each. Deliberately over-sampling negative comments for annotation would address the one class both fine-tuned models still handle poorly, since the limitation is 21 and 74 training examples rather than anything about the method. And collecting reply threads would capture the conversations that top-level comments only hint at, and would test directly whether the low negative shares reported here survive contact with the parts of the discussion this study could not see.

\end{document}